\documentclass[conference]{IEEEtran}
\IEEEoverridecommandlockouts
\usepackage{cite}
\usepackage{amsmath,amssymb,amsfonts}
\usepackage{algorithm}
\usepackage{textcomp}
\usepackage{algorithmic}
\usepackage{xcolor}
\usepackage{url}
\usepackage{float}
\usepackage{amssymb}
\usepackage{multirow}
\usepackage{graphicx}
\usepackage{subcaption}
\usepackage[utf8]{inputenc}
\usepackage[export]{adjustbox}
\usepackage{wrapfig}
\usepackage{tabularx}
\usepackage[font=small,skip=5pt]{caption}
\usepackage{array}
\newcolumntype{?}{!{\vrule width 1pt}}
\usepackage{enumitem}
\makeatletter
\def\hlinewd#1{%
\noalign{\ifnum0=`}\fi\hrule \@height #1 %
\futurelet\reserved@a\@xhline}
\makeatother

\def\BibTeX{{\rm B\kern-.05em{\sc i\kern-.025em b}\kern-.08em
    T\kern-.1667em\lower.7ex\hbox{E}\kern-.125emX}}

\newcommand{\tnew}[2][c]{%
	\begin{tabular}[#1]{@{}c@{}}#2\end{tabular}}

\begin{document}
\renewcommand\citepunct{, }
\title{
Deterministic vs. Non Deterministic Finite Automata in Automata Processing
}

\author{\IEEEauthorblockN{Farzana Ahmed Siddique, Tommy James Tracy II, Nathan Brunelle, Kevin Skadron\\University of Virginia\\\{farzana, tjt7a, njb2b, skadron\}@virginia.edu}
}
\maketitle
\thispagestyle{plain}
\pagestyle{plain}

\begin{abstract}
Linear-time pattern matching engines have seen promising results using Finite Automata (FA) as their computation model. Among different FA variants, deterministic (DFA) and non-deterministic (NFA) are the most commonly used computation models for FA-based pattern matching engines. Moreover, NFA is the prevalent model in pattern matching engines on spatial architectures. The reasons are: i) DFA size, as in \#states, can be exponential compared to equivalent NFA, ii) DFA cannot exploit the massive parallelism available on spatial architectures. This paper performs an empirical study on the \#state of minimized DFA and optimized NFA across a diverse set of real-world benchmarks and shows that \textit{if distinct DFAs are generated for distinct patterns, \#states of minimized DFA are typically equal to their equivalent optimized NFA}. However, NFA is more robust in maintaining the low \#states for some benchmarks. Thus, \textit{the choice of NFA vs. DFA for spatial architecture is less important than the need to generate distinct DFAs for each pattern and support these distinct DFAs' parallel processing}. Finally, this paper presents a throughput study for von Neumann's architecture-based (CPU) vs. spatial architecture-based (FPGA) automata processing engines. The study shows that, \textit{based on the workload, neither CPU-based automata processing engine nor FPGA-based automata processing engine is the clear winner}. If \#patterns matched per workload increases, the CPU-based automata processing engine's throughput decreases. On the other hand, the FPGA-based automata processing engine lacks the memory spilling option; hence, it fails to accommodate an entire automata if it does not fit into FPGA's logic fabric. In the best-case scenario, the CPU has a 4.5x speedup over the FPGA, while for some benchmarks, the FPGA has a 32,530x speedup over the CPU.
\end{abstract}


\section{Introduction}
\label{sec:intro}
Pattern matching is essential in many applications, including genomics, virus detection, network intrusion detection, and machine learning techniques such as random forest. Most of these applications are latency-sensitive. If the pattern matching rate is not fast enough, it acts as a performance bottleneck for those applications. Finite Automata (FA) have proven to be a great computation model for linear time pattern matching \cite{dlugosch2014efficient, xie2017reapr, sidhu2001fast, hieu2013memory, rahimi2020grapefruit}. For example, Schulz et al. used FA to accelerate the pattern matching for genomics applications \cite{ schulz2002fast}. On the other hand, network devices such as the Cisco family of security appliances and networking software such as Snort \cite{roesch1999snort} use regular expressions to represent safe payload patterns; prior works have used FA to accelerate regexes \cite{xie2017reapr, dlugosch2014efficient}.

FA has different variants. The most commonly used FA variants in pattern matching engines are 1. Deterministic Finite Automata (DFA) and 2. Non-deterministic Finite Automata (NFA). Even though both of the variants are functionally equivalent, their structure (state count \& transition) differs \cite{hopcroft2001introduction}. As a result, while designing an FA-based pattern matching engine, the underlying hardware architecture influences the selection between NFA and DFA. NFA is less preferred for von Neumann architecture because the possibility of multiple active states leads to multiple random memory lookups, and high memory latency is a performance bottleneck in von Neumann's architecture \cite{wulf1995hitting}. Micron's Automata Processor \cite{wang2016overview} and spatial architectures, such as Field Programmable Gate Arrays (FPGA), offer workarounds to hide the high memory latency. For these architectures, NFA is a popular choice over DFA \cite{xie2017reapr, rahimi2020grapefruit, tracy2016towards} for several reasons. The crucial one is if an NFA has $n$ states, its equivalent DFA can have $2^n$ states \cite{rabin1959finite,hopcroft2001introduction}. For instance, Yu et al. \cite{yu2006fast} stated that 20\% of the Snort \cite{roesch1999snort} DFAs suffer from exponential state growth relative to the corresponding NFAs. Moreover, prior works show that generating one DFA from a set of patterns could result in an enormous time and space overhead \cite{yu2006fast, becchi2009evaluating}. 

There are algorithms \cite{hopcroft1971n, brzozowski1962canonical} to generate minimized DFAs for which state count is smaller than the expected exponential growth (although exponential growth remains a worst-case for arbitrary automata). Tabakov et al. \cite{tabakov2005experimental} performed a study on synthetic automata and illustrated that based on the automata structure, DFA states exhibit a growth rate of the order polynomial to sub-exponential compared to the NFA, and in some cases, minimized DFAs have a smaller state count than the equivalent non-optimized NFA. Motivated by their study, this paper fills a very important gap in automata processing work. Our paper demonstrates the state count analysis of the minimized DFA and its equivalent optimized NFA using a broad set of real-world benchmarks. The analysis shows that for the real-world benchmarks, the DFA state count exhibits linear or polynomial growth. Therefore, DFAs can be considered a viable alternative to any FA-based pattern matching engine that uses NFA as the computation model. Please note that unlike prior work \cite{tabakov2005experimental}, NFAs used in this analysis are optimized using a heuristics-based approach.  

Another reason for choosing NFA over DFA while designing FA-based pattern matching engines on FPGAs is that NFAs may activate multiple states per symbol cycle, which exploits FPGAs' massive parallelism \cite{paxson1999bro, liu2011fast, tracy2016towards, xie2017reapr}. However, this paper shows that DFAs can also exploit the parallel architecture of the FPGAs if, instead of creating a single large DFA for all the patterns of an application, separate DFAs are generated for distinct patterns, and these DFAs are processed in parallel (section \ref{sec:grpfrt}).

Finally, this paper presents another interesting finding: the state-of-the-art CPU-based automata processing engine, Hyperscan \cite{wang2019hyperscan} sometimes outperforms the state-of-the-art FPGA-based automata processing engine, Grapefruit, in terms of throughput (pattern matching rate). Even though FPGAs offer massive parallelism, FPGAs' clock rate is considerably lower than CPUs. On top of that, applications face performance degradation if the computation model does not fit into the CPU cache. Prior works show that in the worst case, the FPGA-based pattern matching engine has equal throughput compared to the CPU-based pattern matching engine \cite{xie2017reapr, rahimi2020grapefruit}. Contrary to that, this paper shows that for some real-world benchmarks, the CPU gets 4.5x speedup over the FPGA (Section \ref{sec:perf_eval}). However, if a workload is such that it matches many distinct patterns, CPU throughput degrades. So, it can be affirmed that neither CPU nor FPGA is a clear winner for the FA-based pattern matching engine. 

In summary, this paper makes the following contributions:
\begin{itemize}
    \item It presents a detailed analysis of the state count of the optimized NFA vs. minimized DFA across various real-world automata benchmarks (section \ref{sec:state_count_analysis}).
    \item It shows that on FPGA, the choice of NFA over DFA is less important than the need to keep individual DFA separate and the ability to support parallel processing of a large number of DFAs (section \ref{sec:grpfrt}).
    \item This paper presents an in-depth performance analysis of the state-of-the-art CPU-based automata processing engine vs. the state-of-the-art FPGA-based automata processing engine (section \ref{sec:perf_eval}).
\end{itemize}

\begin{figure*}[h]
\captionsetup[subfigure]{aboveskip=0pt,belowskip=5pt}
\begin{subfigure}{0.33\textwidth}
\includegraphics[clip,  trim={0cm 0cm 0cm 0cm},scale=0.33]{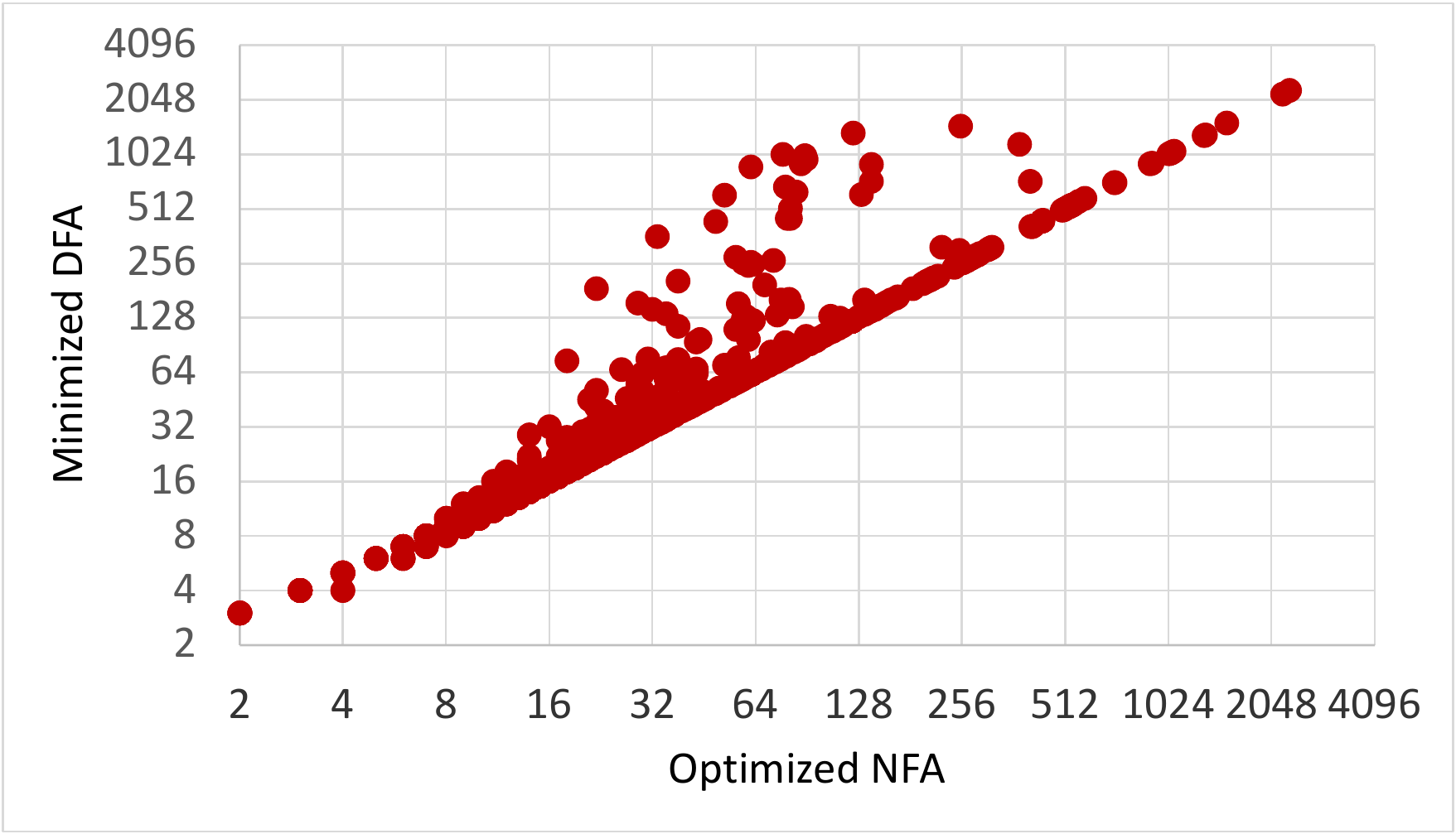}
\caption{Snort}
\label{fig:snort_st}
\end{subfigure}
\begin{subfigure}{0.33\textwidth}
\includegraphics[clip,  trim={0cm 0cm 0cm 0cm},scale=0.33]{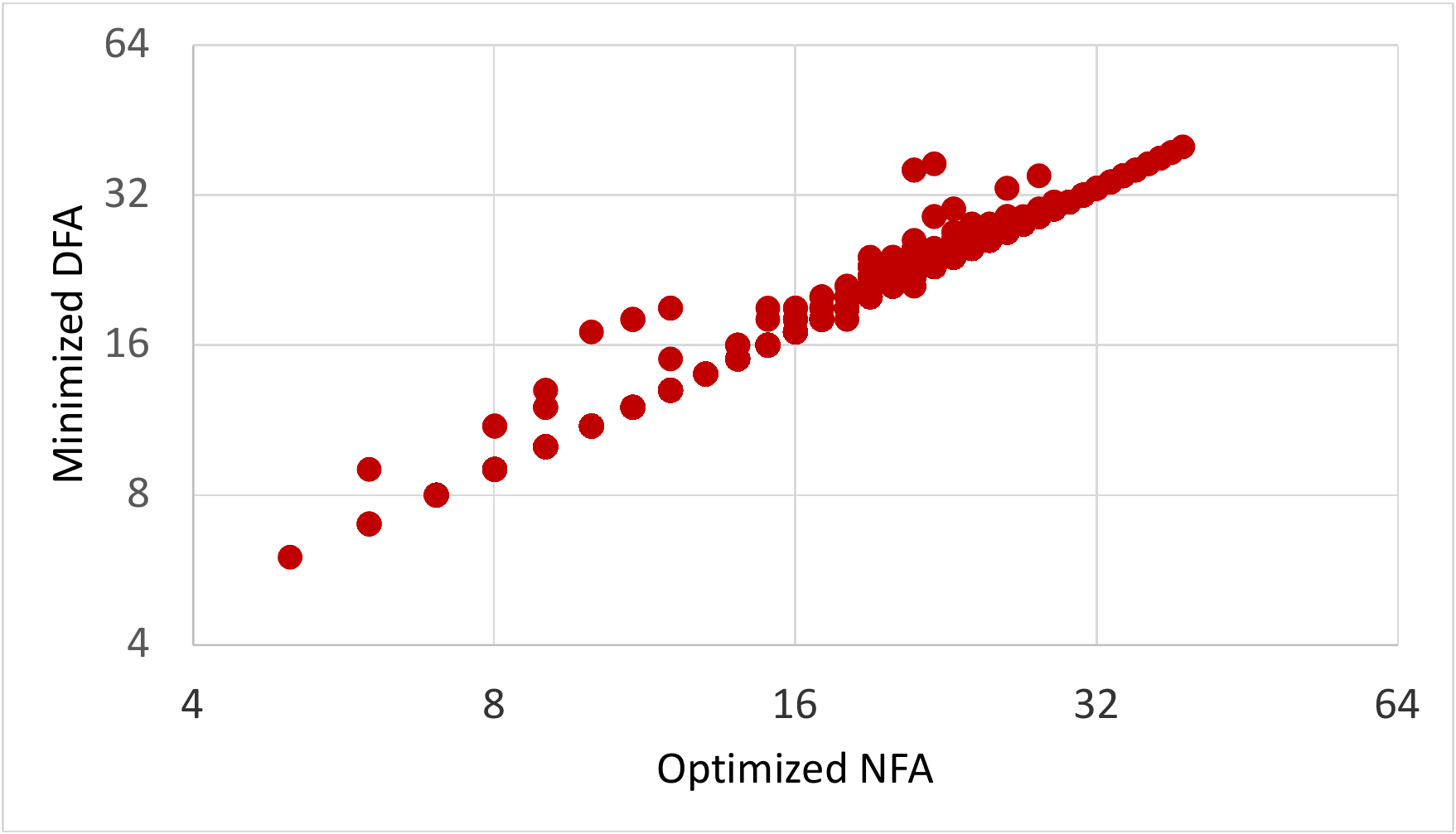}
\caption{Brill}
\label{fig:brill_st}
\end{subfigure}
\begin{subfigure}{0.33\textwidth}
\includegraphics[clip,  trim={0cm 0cm 0cm 0cm},scale=0.33]{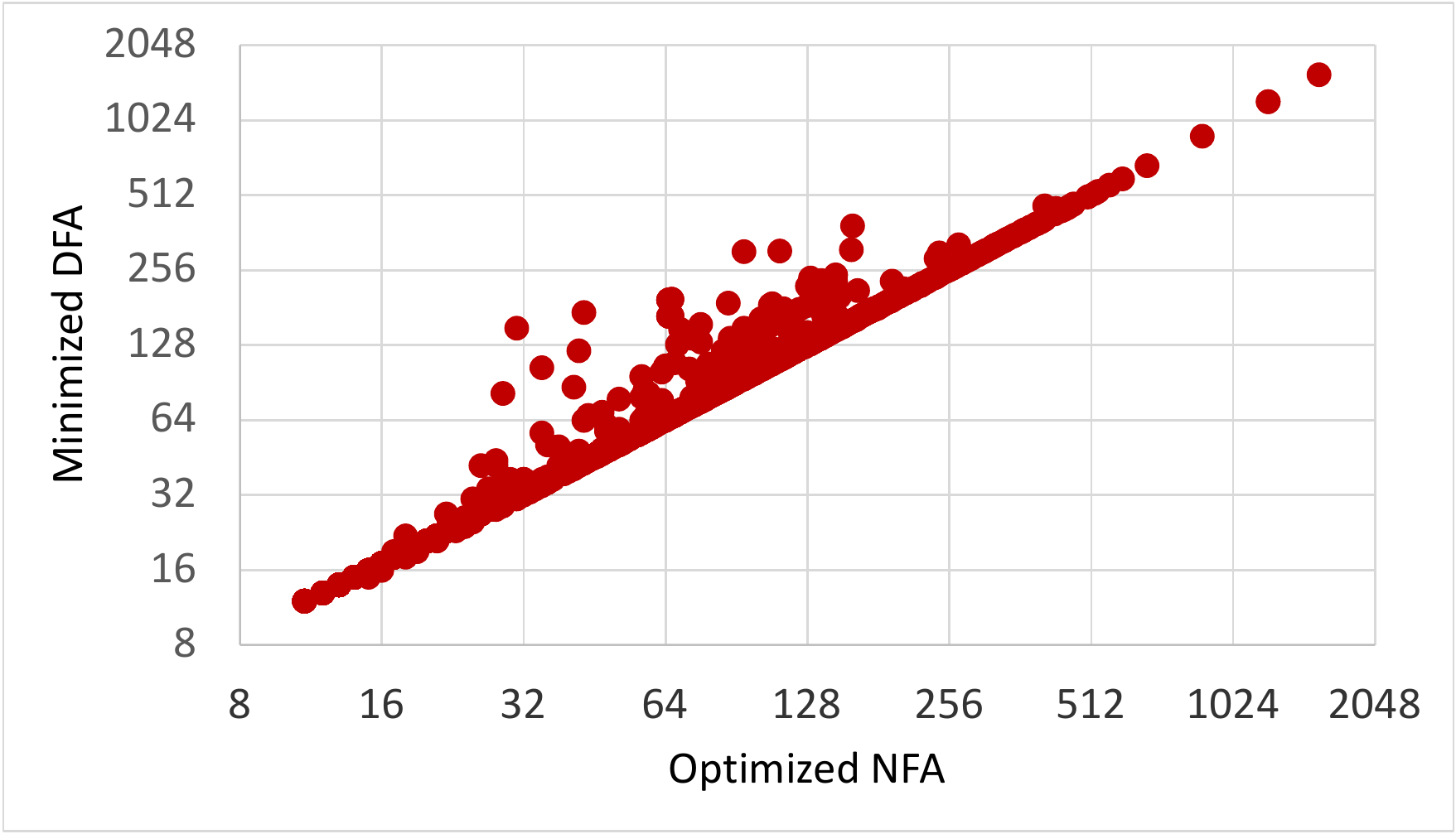}
\caption{ClamAV}
\label{fig:clam_st}
\end{subfigure}
\begin{subfigure}{0.33\textwidth}
\includegraphics[clip,  trim={0cm 0cm 0cm 0cm},scale=0.33]{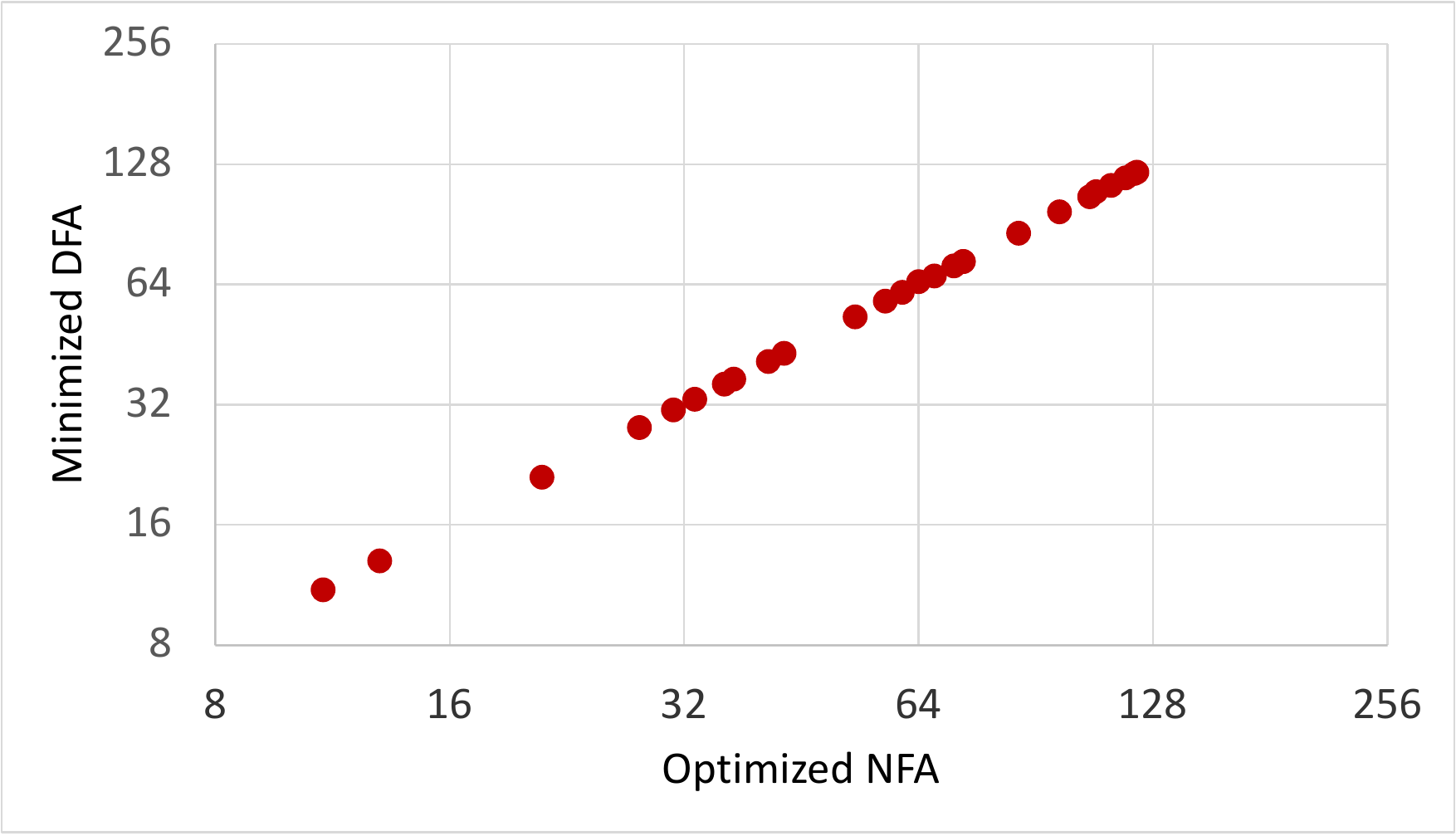}
\caption{Hamming}
\label{fig:ham_st}
\end{subfigure}
\begin{subfigure}{0.33\textwidth}
\includegraphics[clip,  trim={0cm 0cm 0cm 0cm},scale=0.33]{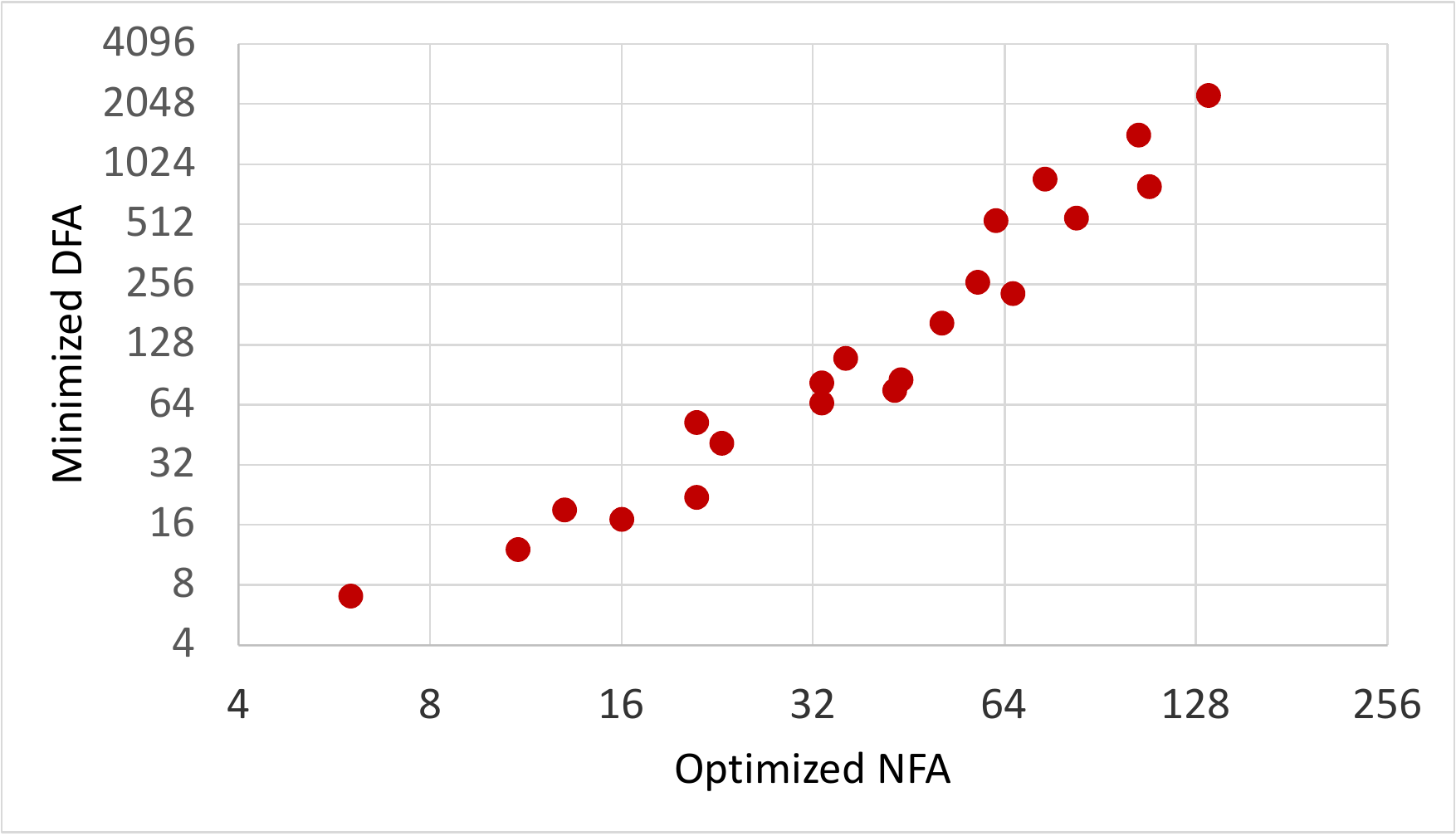}
\caption{Levenshtein}
\label{fig:lev_st}
\end{subfigure}
\begin{subfigure}{0.33\textwidth}
\includegraphics[clip,  trim={0cm 0cm 0cm 0cm},scale=0.33]{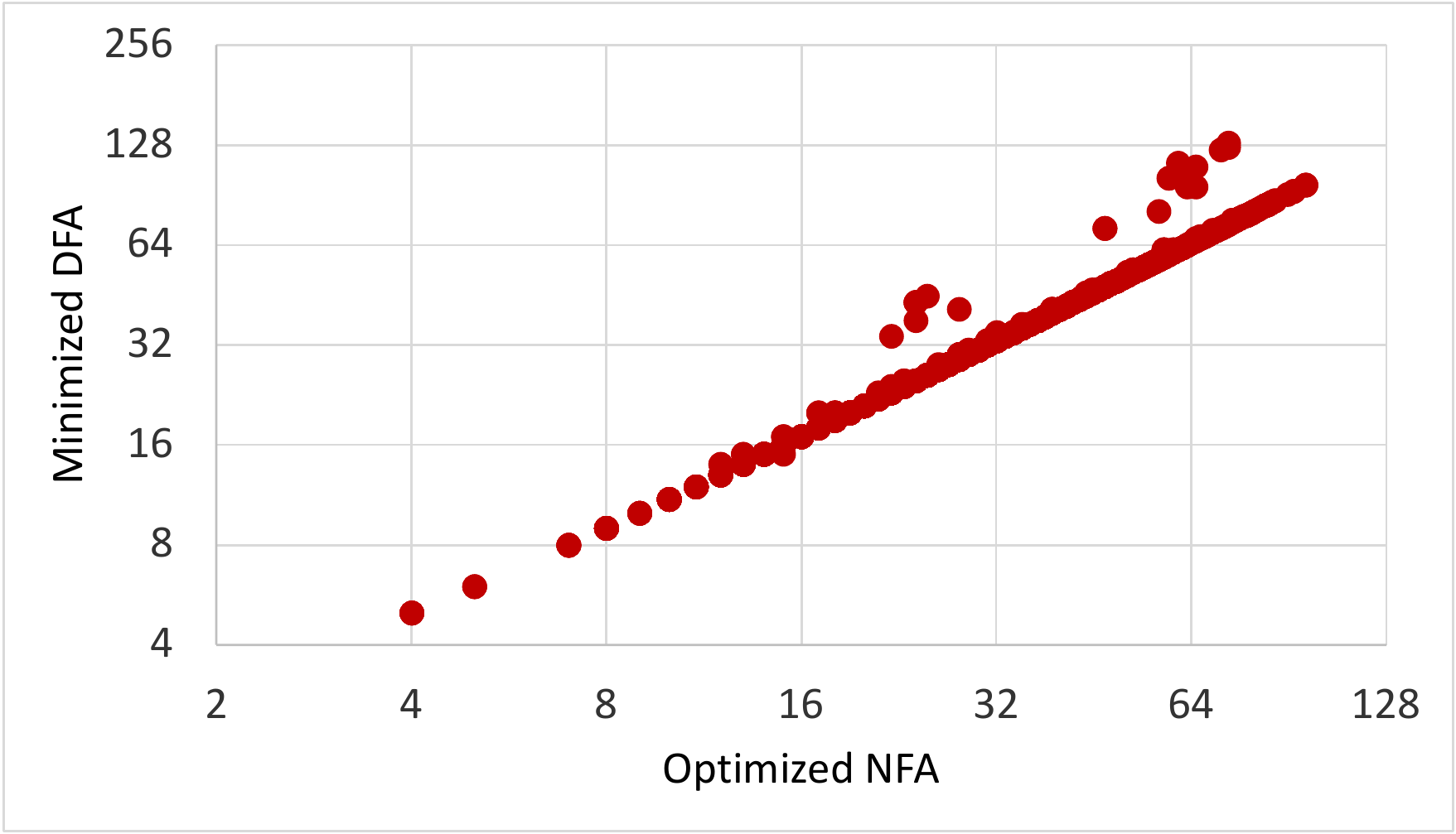}
\caption{DotStar}
\label{fig:dot_st}
\end{subfigure}
\caption{State count comparison between optimized NFA and minimized DFA by generating distinct DFA/NFA for each individual pattern (shown in log-log scale). For Snort, Brill, ClamAV, and DotStar a few of the DFA's states have polynomial growth rate compared to the NFA states. Hamming DFA state count is always equal to its equivalent NFA's state count. DFA shows polynomial growth in state count compared to it's equivalent NFA for Levenshtein.}
\label{fig:statecomp}
\end{figure*}

\begin{figure*}[ht]
\begin{subfigure}{0.33\textwidth}
\includegraphics[clip,  trim={0cm 0cm 0cm 0cm},scale=0.33]{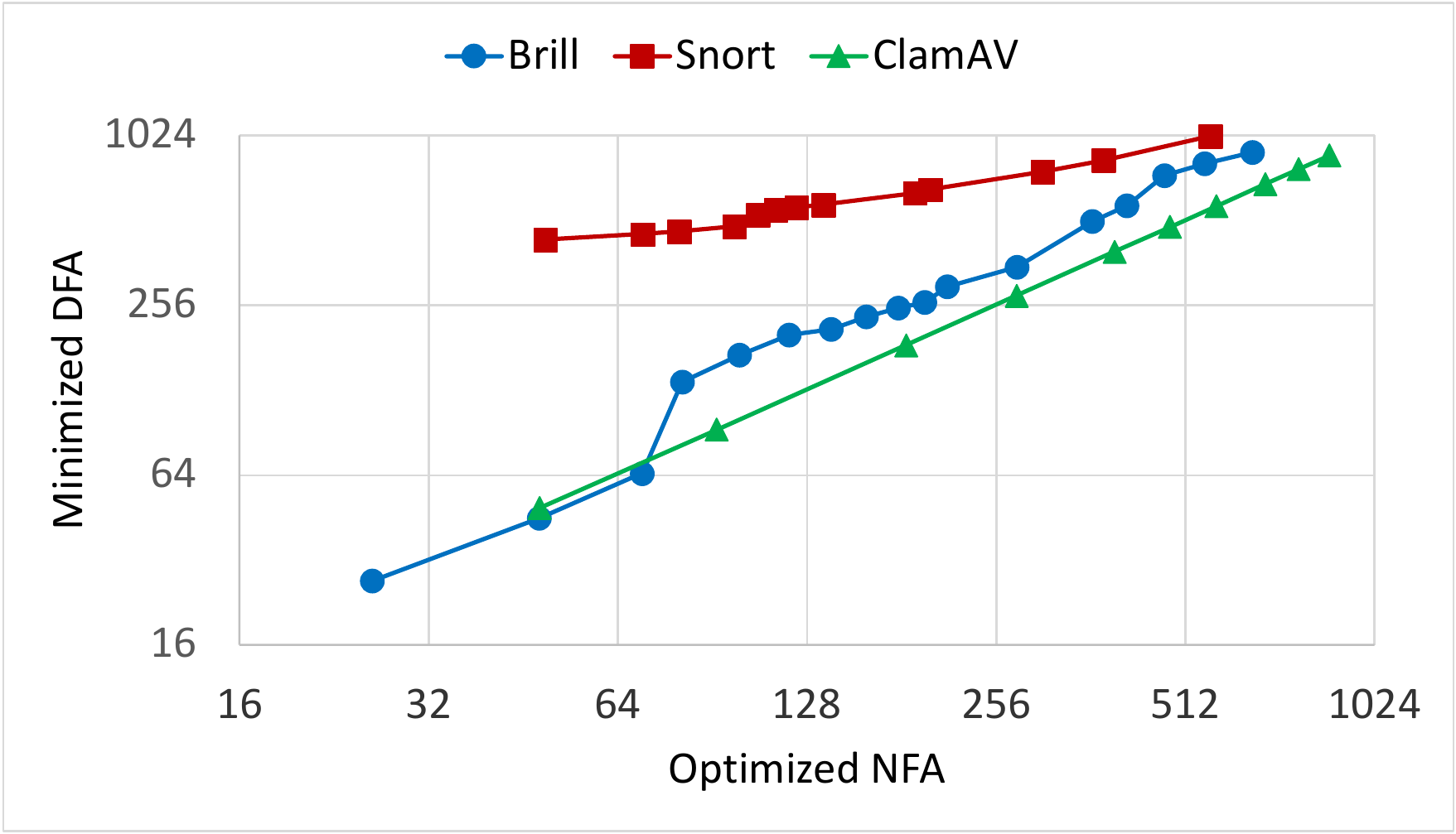}
\caption{Regex Benchmark}
\label{fig:reg_multi}
\end{subfigure}
\begin{subfigure}{0.33\textwidth}
\includegraphics[clip,  trim={0cm 0cm 0cm 0cm},scale=0.33]{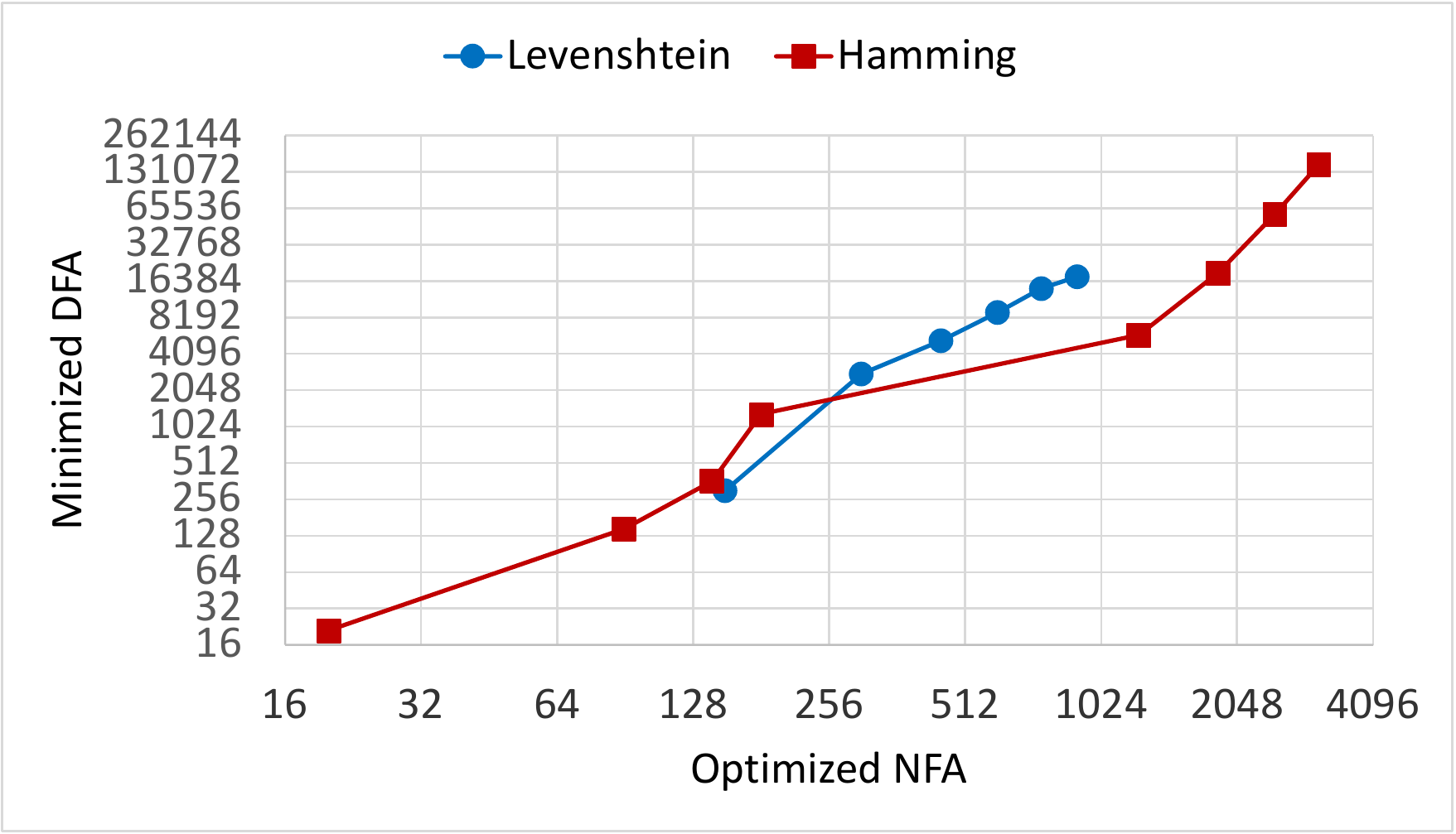}
\caption{Mesh Benchmark}
\label{fig:ham_st_multi}
\end{subfigure}
\begin{subfigure}{0.33\textwidth}
\includegraphics[clip,  trim={0cm 0cm 0cm 0cm},scale=0.33]{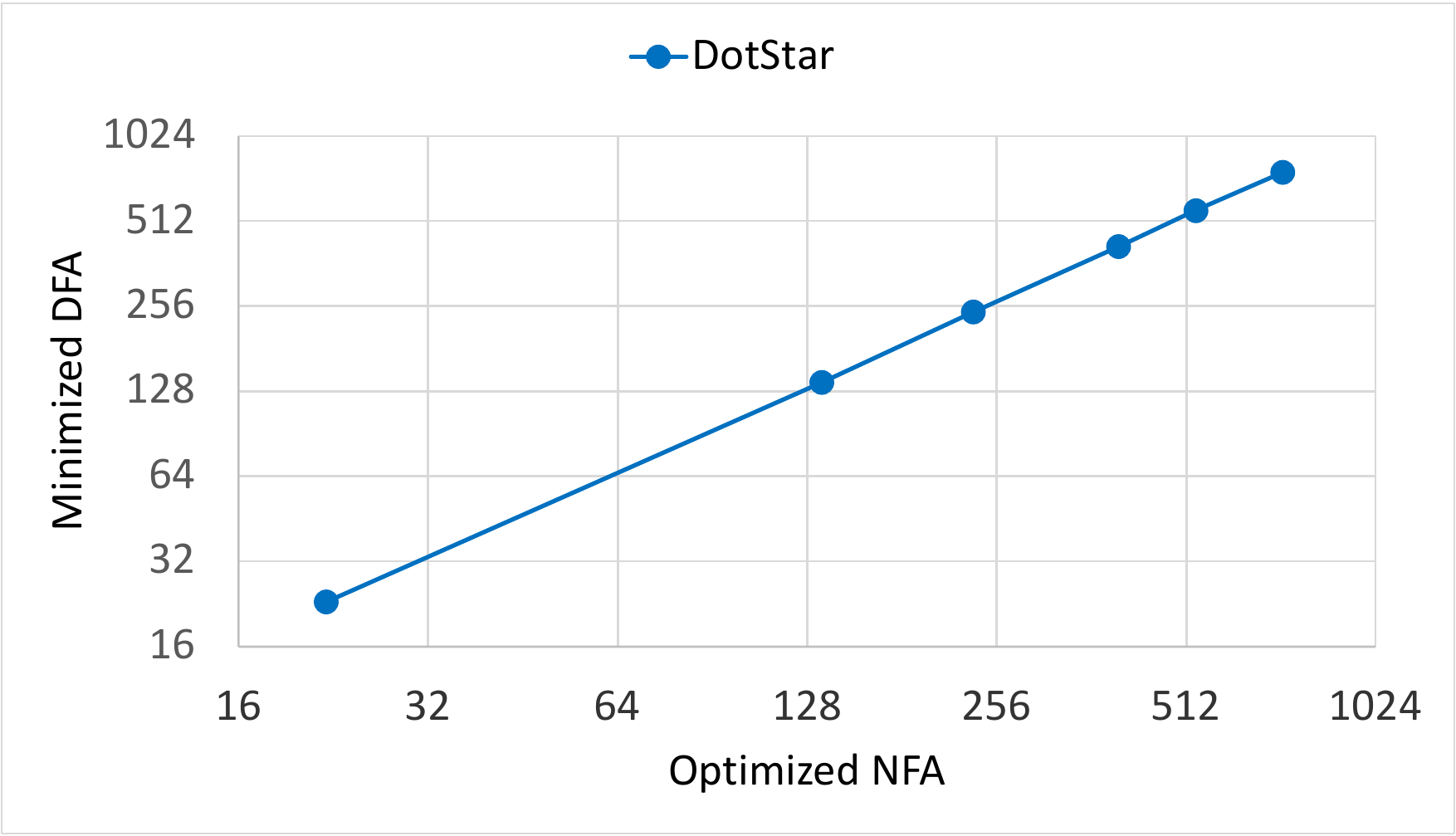}
\caption{Synthetic Benchmark}
\label{fig:clam_st_multi}
\end{subfigure}
\caption{State count comparison between optimized NFA and minimized DFA by merging multiple patterns into a single automaton (shown in log-log scale). Regex-based benchmarks show linear growth-rate in state count for minimized DFA compared to its equivalent optimized NFA. Minimized DFAs for Mesh-Benchmarks exhibit exponential growth-rate while for synthetic benchmark, they have equal number of states as their equivalent NFA.}
\label{fig:statecompmulti}
\end{figure*}
\section{Related Work}
The state count of the NFA and DFA has been studied previously from the perspective of network-intrusion detection, and synthetic benchmarks \cite{becchi2007hybrid, yu2006fast, kumar2007curing}. The studies on the network-intrusion detection benchmarks showed that generating a DFA for a specified pattern set can exponentially increase the DFA state count. They demonstrate that regular expressions containing constraints such as dot-star (pq.*rs) and repeated sub-patterns (xyz.\{100\}abc) induce exponential growth in DFA states. Contrary to the prior works, this paper performs state count analysis for a diverse state of real-world benchmarks and shows that if separate DFAs are generated for each pattern, the DFA state count is equal to the equivalent optimized NFA. However, in the worst-case DFA state count shows a polynomial growth rate.

Tabakov et al. \cite{tabakov2005experimental} evaluated Hopcroft's \cite{hopcroft1971n} and Brzozowski's \cite{brzozowski1962canonical} DFA minimization algorithm using a synthetic automata type called random automata. Their analysis illustrates that the state count of a DFA depends on the automata graph's transition density\footnote{For each symbol ratio of the total number of transitions to the total number of states. Details can be found at \cite{tabakov2005experimental}.}. For smaller transition densities, the DFA state grows super-polynomially but sub-exponentially. Additionally, their result shows that minimized DFA has a smaller state count than the equivalent NFAs with greater transition densities. One limitation of this analysis is that the NFAs used are not optimized. Even though NFA minimization is a PSPACE-complete problem \cite{jiang1993minimal}, a heuristics-based approach can generate optimized NFA, as shown in this paper.

Since Floyd et al. \cite{floyd1980compilation} introduced hardware implementation of NFA, there has been a substantial amount of work on high-performance automata processors using either DFA \cite{hieu2013memory, yu2006fast, kumar2006algorithms} or NFA \cite{sidhu2001fast, xie2017reapr, yang2011high, dlugosch2014efficient} as a computation model. The state-of-the-art CPU-based automata processing engine Hyperscan \cite{wang2019hyperscan} uses regex decomposition to separate the pattern into disjoint strings and FA. Moreover, to process FA, Hyperscan exploits SIMD vector operation to perform the maximum possible state transition for each memory access. Hyperscan encounters throughput degradation due to the CPU's high memory latency if the working set size does not fit into the CPU cache. Grapefruit \cite{rahimi2020grapefruit}, a state-of-the-art FPGA-based automata processing engine, maps a cluster of automata into the FPGA logic fabric. Even though this approach helps in accelerating Grapefruit's throughput, if the automata graph does not fit into the FPGA logic fabric, Grapefruit fails to process it. This paper does not discuss GPU-based automata processing engine because prior works \cite{DFAGE, liu2020gpus} have presented that GPU is not promising for processing FA.

Xie et al. \cite{xie2017reapr} examined throughput comparison between the state-of-the-art CPU-based automata processing engine, Hyperscan \cite{wang2019hyperscan} and an FPGA-based automata processing engine, REAPR. Their analysis shows that in the worst case, REAPR performs the same as Hyperscan, whereas, in the best case, REAPR has a 2,000x speedup over Hyperscan. However, this paper shows that, based on the workload and the benchmark, neither Hyperscan nor Grapefruit is a clear winner for the FA-based pattern matching engine.

Becchi et al. \cite{becchi2009evaluating} presented a performance evaluation of a network processor and a general-purpose processor using NFA, DFA, and hybrid-FA. Hybrid-FA is a variant of automata formed by halting the NFA$\rightarrow$DFA conversion in places where DFA blowup occurs. The evaluation at \cite{becchi2009evaluating} shows that hybrid-FA has the overall best throughput across the network and general-purpose processors. The reason is that the DFA part of the hybrid-FA is less susceptible to high memory latency, and the NFA part of the hybrid-FA keeps a check on the state explosion issue of the DFA. This performance evaluation is limited to von Neumann architecture, whereas this paper does the performance evaluation for von Neumann vs. spatial architecture.
\section{NFA vs. DFA State Count Analysis}
\label{sec:state_count_analysis}
By definition, if an NFA requires $n$ states to express a regular language $L$ or pattern $P$, the equivalent DFA will require at most $2^n$ states \cite{hopcroft2001introduction}. As a result, DFAs are disregarded for designing FA-based pattern matching engines on FPGA \cite{becchi2007hybrid, xie2017reapr}. The state count study of DFA vs. NFA across diverse real-world benchmarks is extremely important for automata processing research. Because it would resolve if, for real-world benchmarks, minimized DFAs' state count exhibits an exponential growth rate compared to its equivalent NFA. The scope of prior works that perform such study is limited to network intrusion detection and synthetic benchmarks \cite{yu2006fast, tabakov2005experimental}. This section presents an empirical analysis of the state count of DFA vs. NFA using a diverse benchmark set taken from the real workload \cite{wadden2018automatazoo}. 

While a DFA representing multiple distinct patterns may be significantly larger \cite{yu2006fast}, a simplified, single pattern DFA can be as small as its equivalent NFA \cite{hopcroft1971n}. Thus, this paper performs state count analysis from two different perspectives. First, the state count for minimized DFA and optimized NFA is analyzed by generating distinct NFAs and DFAs for each pattern (Fig \ref{fig:statecomp}). Second, the state count is compared by incrementally adding patterns to a single NFA or DFA (Fig \ref{fig:statecompmulti}). These two analyses show that while generating individual DFA for each pattern, DFA's state count is typically equal to the state count of the NFA. However, few DFAs exhibit polynomial growth in their state count in the worst case. If multiple patterns are merged into a single DFA, the DFA state count may show exponential growth compared to the equivalent NFA.
\subsubsection*{Experimental Setup}
AutomataZoo \cite{wadden2018automatazoo} benchmark suite has been used for this experiment. Each benchmark of AutomataZoo comes with a single NFA, and that NFA contains all the patterns for that particular benchmark. However, the state count analysis needs distinct NFAs representing distinct patterns, so VASim's \cite{wadden2016vasim} \emph{compute connected component} option is used to divide individual patterns into separate NFA graphs. Then using Brzozowski's DFA minimization algorithm, minimized DFA is generated for each NFA. NFA minimization is a PSPACE-complete problem \cite{jiang1993minimal}. So, a heuristic-based approach is used to generate optimized NFA: if multiple states have \textit{equivalent outgoing transitions} (for all the symbols, they transition into the same states), merge those states into one state. All these steps compare the state count of NFA vs. DFA for distinct patterns. An extra step is needed for the comparison where multiple patterns are merged into a single NFA/DFA graph, incrementally add patterns into an NFA and create corresponding minimized DFA and optimized NFA using the above approach.

\subsubsection*{Evaluation}
Figure \ref{fig:statecomp} \& \ref{fig:statecompmulti} shows the summary of optimized NFA vs. minimized DFA state count analysis.
\begin{enumerate}
    \item Among the three regex-based benchmarks,
    (Fig \ref{fig:snort_st}, \ref{fig:brill_st}, \ref{fig:clam_st}) most of the minimized DFAs have an equal number of states as their NFAs. However, a few minimized DFAs have a polynomial growth rate in state count compared to the equivalent NFA. When patterns are merged, these regex benchmarks exhibit linear growth rate in state count for minimized DFA (Fig \ref{fig:reg_multi}).
    \item There are two mesh automata benchmarks in AutomataZoo. Between them, Hamming benchmark's minimized DFAs have an equal number of states as their equivalent optimized NFAs (Fig \ref{fig:ham_st}). Whereas the Levenshtein benchmark's minimized DFA state count has a polynomial growth compared to the equivalent optimized NFA (Fig \ref{fig:lev_st}). On the other hand, if patterns are merged, the state count grows exponentially for minimized DFA compared to the equivalent NFA.
    \item Synthetic benchmarks such as DotStar show a linear growth in minimized DFA state count except for some polynomial growth (Fig \ref{fig:dot_st}). The DFA state count shows linear growth when patterns are merged for this benchmark.
    \item All the patterns for the Random Forest \cite{tracy2016towards} benchmark have an equal number of states for the NFAs. Moreover, the state count of DFA is the same as the NFA. As there is only one data point for this benchmark, it is not included in Fig \ref{fig:statecomp}.
\end{enumerate}

\begin{table}
    \caption{NFA vs. DFA Performance Analysis on FPGA.}
    \label{tab:hw_exp}
    \centering
    \renewcommand{\arraystretch}{1.1}
    \setlength\tabcolsep{1pt}
    \scriptsize
    \begin{tabular}{?c|c|c|c|c|c|c|c?}
        \hlinewd{1pt}
         Benchmark & \tnew{\#Patterns} & \tnew{FSA} & \tnew{\#States \\ (k)} & \tnew{LUTs \\ (k)} & \tnew{FFs\\(k)} & \tnew{Frequency\\(MHz)} & \tnew{Throughput\\(Gbps)}\\
         \hlinewd{1pt}
         \multirow{2}{*}{Brill} & \multirow{2}{*}{5,946} & NFA & 116 & 151 & 112 & 500 & 4 \\
         \cline{3-8}
         & &mDFA & 126 & 163 & 115 & 345 & 2.8\\
         \hlinewd{1pt}
         \multirow{2}{*}{Snort} & \multirow{2}{*}{2,348} & NFA & 143 & 163 & 135 & 200 & 1.6\\
         \cline{3-8}
         & & mDFA & 196 & 277 & 168 & 103 & 0.8\\
         \hlinewd{1pt}
         \multirow{2}{*}{ClamAV} & \multirow{2}{*}{14,300} & NFA & 848 & 1,047 & 848 & 172 & 1.4 \\
         \cline{3-8}
         & & mDFA & 851 & 1,058 & 849 & 164 & 1.3 \\
         \hlinewd{1pt}
         \multirow{2}{*}{YARA} & \multirow{2}{*}{2,620} & NFA & 115 & 136 & 115 & 194 & 1.5 \\
         \cline{3-8}
         & & mDFA & 120 & 150 & 118 & 333.33 & 2.7\\
         \hlinewd{1pt}
         \multirow{2}{*}{Hamming} & \multirow{2}{*}{1,000} & NFA & 108 & 56 & 64 & 556 & 4.4 \\
         \cline{3-8}
         & & mDFA & 108 & 67 & 63 & 556 & 4.4\\
         \hlinewd{1pt}
         \multirow{2}{*}{Levenshtein} & \multirow{2}{*}{1,000} & NFA & 106 & 125 & 103 & 263 & 2.1 \\
         \cline{3-8}
         & & mDFA & 735 & 783 & 528 & 222 & 1.8\\
         \hlinewd{1pt}
         \multirow{2}{*}{\tnew{Random\\Forest}} & \multirow{2}{*}{8,000} & NFA & 248 & 407 & 240 & 303 & 2.4 \\
         \cline{3-8}
         & & mDFA & 248 & 409 & 240 & 257 & 2.1\\
         \hlinewd{1pt}
         \multirow{2}{*}{\tnew{Entity\\Resolution}} & \multirow{2}{*}{10,000} & NFA & 413 & 465 & 260 & 204 & 1.6 \\
         \cline{3-8}
         & & mDFA & 422 & 541 & 275 & 250 & 2.0\\
         \hlinewd{1pt}
         \multirow{2}{*}{FileCarving} & \multirow{2}{*}{9} & NFA & 0.179 & 0.187 & 0.122 & 666.67 & 5.3 \\
         \cline{3-8}
         & & mDFA & 0.184 & 0.211 & 0.128 & 666.67 & 5.3\\
         \hlinewd{1pt}
    \end{tabular}
\end{table}
\section{NFA vs. DFA Performance Analysis on FPGA}
\label{sec:grpfrt}
Prior works on FPGA-based automata processing engine \cite{xie2017reapr, rahimi2020grapefruit, sidhu2001fast} selected NFA as their computation model because NFA enables processing multiple states simultaneously, which exploits FPGA's parallel execution capability. Prior works also pointed out that DFA conversion and processing are expensive as DFA might suffer from exponential state growth \cite{becchi2007hybrid}. In contrast, Section \ref{sec:state_count_analysis} of this paper shows that the DFA conversion will not be as expensive if distinct DFA is generated per individual pattern. This section shows that DFAs can also exploit parallelism offered by FPGA if distinct DFAs are processed in parallel (table \ref{tab:hw_exp}). 
\subsubsection*{Experimental Setup}
For this analysis, state-of-the-art FPGA-based automata processing engine, Grapefruit \cite{rahimi2020grapefruit} is used to generate the Hardware Description Language (HDL) for the synthesis, and the FPGA throughput for NFA/DFA is calculated using the maximum frequency reported after finishing the synthesis. The DFAs and NFAs used in this experiment are all optimized.
\subsubsection*{Evaluation}
Table \ref{tab:hw_exp} presents the detailed result of the experiment. On the FPGA, DFA has 1.8x and 1.3x speedup for YARA and Entity Resolution benchmarks, respectively. Further investigation shows that NFA graphs have a higher fan-out or node degree for these two benchmarks than the DFA. High fan-out or node degree works as a performance bottleneck for FPGA processing. FPGA has identical throughput for ClamAV, Hamming, and Random Forest benchmarks for both the NFA and DFA. For the rest of the benchmarks, FPGA has 2x-1.1x speedup while processing NFA instead of DFA.
\section{CPU vs. FPGA Performance Analysis} 
\label{sec:perf_eval}
Researchers designing high-performance FA-based pattern matching engines prefer FPGA over CPU because i) CPU has limited parallelism, while FPGA offers massive parallelism, ii) CPU faces high memory latency, but FPGA does not as automata states can be mapped into the logic fabric \cite{xie2017reapr, rahimi2020grapefruit}. This section shows that the choice of CPU vs. FPGA is not trivial.

High memory latency negatively affects the CPU's throughput if the automata graph does not fit into the cache. On the other hand, FPGA has a significantly lower clock rate compared to the CPU. For example, in practice, FPGA's achieve 200-500MHz, while high-end CPUs are 2-3 GHz or more. Besides, if the entire benchmark do not fit into the FPGA logic fabric, FPGA needs an optimal memory spilling technique to accommodate all the patterns of the benchmark. For example, even though Grapefruit \cite{rahimi2020grapefruit} is the current state-of-the-art FPGA-based pattern matching engine, it fails to process the entire ClamAV benchmark (33k patterns) because it does not have a memory spilling option. Instead, it breaks the application into chunks and processes those in multiple passes over the input. On the other hand, the state-of-the-art CPU-based pattern matching engine, Hyperscan \cite{wang2019hyperscan}, can, in fact, process the entire ClamAV benchmark. Not only that, Hyperscan has 4.5x speedup over Grapefruit, even though Grapefruit was processing only 43.3\% of the ClamAV benchmark. Having said that, Grapefruit has 32,530x speedup over Hyperscan for the APPRNG benchmark. So, it can safely be asserted that neither Grapefruit nor Hyperscan is the clear winner here.
 \begin{table}
    \caption{Hyperscan vs. Grapefruit Throughput Comparison}
    \label{tab:hps_grp}
    \centering
    \renewcommand{\arraystretch}{1.3}
    \setlength\tabcolsep{6pt}
    \begin{tabular}{?c|c|c?}
        \hlinewd{1pt}
         Benchmark & \tnew{Hyperscan\\ Throughput\\(Gbps)} & \tnew{Grapefruit\\ Throughput\\(Gbps)}\\
         \hlinewd{1pt}
         Brill & 0.007 & 4 \\
         \hlinewd{1pt}
         Snort & 0.359 & 1.6 \\
         \hlinewd{1pt}
         SeqMatch & 0.020 & 2.3\\
         \hlinewd{1pt}
         Levenshtein & 0.0002 & 2.1\\
         \hlinewd{1pt}
         Hamming & 0.0003 & 4.4 \\
         \hlinewd{1pt}
         Protomata & 0.016 & 2.9 \\
         \hlinewd{1pt}
         Random Forest & 0.005 & 2.4 \\
         \hlinewd{1pt}
         Entity Resolution & 0.014 & 1.6 \\
         \hlinewd{1pt}
         APPRNG & 0.000083 & 2.7 \\
         \hlinewd{1pt}
         ClamAV & 6.245 & 1.4 \\
         \hlinewd{1pt}
         YARA & 6.877 & 1.5\\
         \hlinewd{1pt}
         FileCarve & 22.475 & 5.3\\
         \hlinewd{1pt}
    \end{tabular}
\end{table}
 
\subsubsection*{Experimental Setup} 
Hyperscan is used for the CPU-based pattern matching engine and Grapefruit as the FPGA-based pattern matching engine. Hyperscan is run on an Intel i7-6700K CPU running at 4 GHz with 32 GB memory. Intel Xeon E5-2686 v4 processor running at 2.3 GHz and 32 GB memory is used as FPGA host computer. Following approach explains the Hyperscan throughput calculation step for each benchmark of the AutomataZoo \cite{wadden2018automatazoo}:
\begin{enumerate}
    \item Compile: Hyperscan takes regular expression and compiles it into a pattern database \cite{hsdb}. However, all the benchmarks in AutomataZoo have their automata expressed either in ANML or MNRL format. As a result, to feed the automata to Hyperscan, I use hscompile's \cite{hscompile} to compile the MNRL file to a pattern database. 
    \item Scan: I use $hsrun$ \cite{hsscan} executable of the hscompile tool to scan the input file of each benchmark, and log the scan time for this process. Then I calculate the throughput using equation \ref{eq:hps_th}.
\end{enumerate}
\begin{equation}
\label{eq:hps_th}
throughput_{Hyperscan}(Gbps) = \frac{Input\_Size(Gb)}{Scan\_Time(sec)}
\end{equation}
Hyperscan uses regex decomposition as one of its throughput optimization techniques. The throughput obtained here is cross-checked by running Hyperscan with the regex version of the benchmarks to confirm that the optimization technique does not fall short if NFAs are directly fed to Hyperscan.

Grapefruit throughput is calculated only for computation kernel, excluding I/O, based on the fact that the FPGA process one input symbol per clock.
\subsubsection*{Evaluation}
\begin{figure}[hb]
\centering
\includegraphics[clip, trim={0cm 0cm 0cm 0cm},scale=0.5]{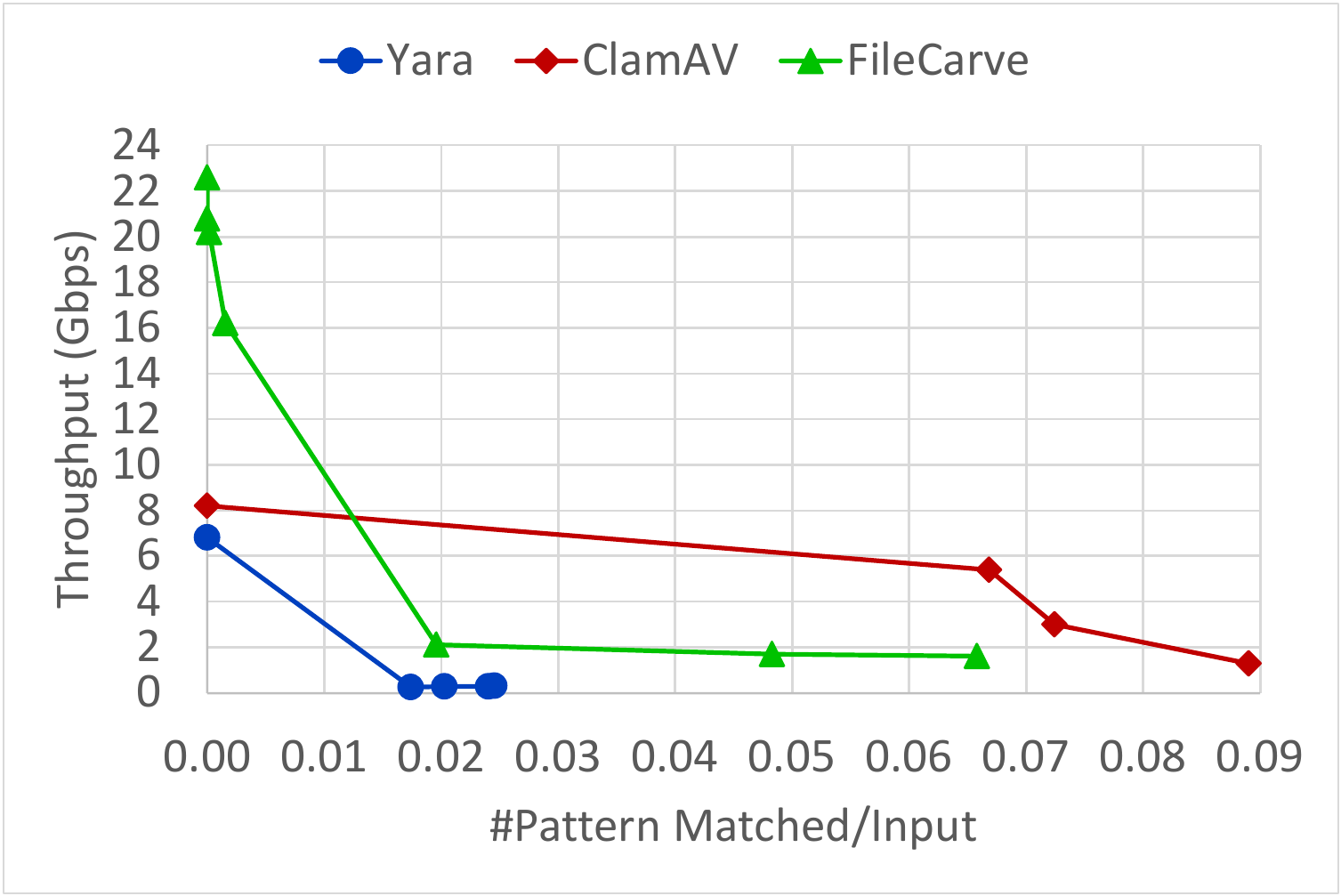}
\label{fig:dfa}
\caption{Hyperscan throughput decreases as \#patterns matched/input increases} 
\label{fig:hpscn_thrpt_sensitivity}
\end{figure}
Table \ref{tab:hps_grp} shows the throughput (Gbps) comparison between the Grapefruit and the Hyperscan. Except for ClamAV, YARA, and FileCarve, Grapefruit has significantly higher throughput (max speedup 32,530x) than Hyperscan. FileCarve is a special case as the benchmark size is very small (only nine patterns). However, further exploration into the characteristic (\#patterns, \#pattern matched per input) of ClamAV and YARA benchmarks show that the \#pattern matched per input is significantly low for these benchmarks, which means only a small portion of the automata graphs are traversed.

This paper hypothesized that Hyperscan throughput is sensitive to \#pattern matched per input. In support of this hypothesis, another empirical study is performed. First, a set of input is generated synthetically for the benchmarks. Next, Hyperscan's throughput is calculated for those benchmarks using the synthetic inputs. Fig \ref{fig:hpscn_thrpt_sensitivity} shows the result for this analysis. It can be seen that, Hyperscan's throughput decreases as the \#pattern matched per input increases. The findings in this section shows that, the CPU might not offer high parallelism or suffer from high memory latency, but the CPU is better than the FPGA if the workload does not cause high memory access. Thus, a high-performance FA-based pattern matching engine may entail a heterogeneous system, where workloads are divided intelligently between the CPU and FPGA.

\section{Conclusion}
Tabakov et al. \cite{tabakov2005experimental} showed that sometimes DFAs can be more compact in state count than their equivalent NFAs. However, their work does not use optimized NFAs, while their DFAs are minimized. Motivated by their work, this paper performed a state count analysis for minimized DFA and heuristically-optimized NFA across a diverse set of real-world automata benchmarks. The analysis focuses on two perspectives: i) distinct DFA and NFA are generated for each pattern, and ii) multiple patterns are merged into a single DFA/NFA. This study shows that if distinct DFAs/NFAs are generated for each pattern, minimized DFAs' state count growth rate is often similar to its equivalent optimized NFA for these benchmarks. However, some DFAs' state counts show exponential growth rates if multiple patterns are merged into a single DFA. Based on this finding, we hypothesized that if individual distinct DFAs are generated for each distinct pattern and are processed in parallel, the FPGA throughput for processing DFA would be similar to the FPGA throughput for processing NFA. Furthermore, an empirical analysis of the NFA vs. DFA throughput presented in this paper supports this hypothesis. Finally, unlike prior work \cite{rahimi2020grapefruit, xie2017reapr}, this paper shows that based on the workload and the benchmark, neither CPU nor FPGA is a clear winner for the FA-based pattern matching engine. Further optimization of our FPGA framework is ongoing.
\section*{Acknowledgement}
This work was supported by Center for Research on Intelligent Storage and Processing in Memory (CRISP) and Semiconductor Research Corporation (SRC).
\bibliographystyle{vancouver}
\bibliography{reference}

\begin{thebibliography}{10}

\bibitem{dlugosch2014efficient}
Dlugosch P, Brown D, Glendenning P, Leventhal M, Noyes H.
\newblock An efficient and scalable semiconductor architecture for parallel
  automata processing.
\newblock IEEE Transactions on Parallel and Distributed Systems.
  2014;25(12):3088-98.

\bibitem{xie2017reapr}
Xie T, Dang V, Wadden J, Skadron K, Stan M.
\newblock REAPR: Reconfigurable engine for automata processing.
\newblock In: 2017 27th International Conference on Field Programmable Logic
  and Applications (FPL). IEEE; 2017. p. 1-8.

\bibitem{sidhu2001fast}
Sidhu R, Prasanna VK.
\newblock Fast regular expression matching using FPGAs.
\newblock In: The 9th Annual IEEE Symposium on Field-Programmable Custom
  Computing Machines (FCCM'01). IEEE; 2001. p. 227-38.

\bibitem{hieu2013memory}
Hieu TT, Tran NT.
\newblock A memory efficient FPGA-based pattern matching engine for stateful
  NIDS.
\newblock In: 2013 Fifth International Conference on Ubiquitous and Future
  Networks (ICUFN). IEEE; 2013. p. 252-7.

\bibitem{rahimi2020grapefruit}
Rahimi R, Sadredini E, Stan M, Skadron K.
\newblock Grapefruit: An Open-Source, Full-Stack, and Customizable Automata
  Processing on FPGAs.
\newblock In: 2020 IEEE 28th Annual International Symposium on
  Field-Programmable Custom Computing Machines (FCCM). IEEE; 2020. p. 138-47.

\bibitem{schulz2002fast}
Schulz KU, Mihov S.
\newblock Fast string correction with Levenshtein automata.
\newblock International Journal on Document Analysis and Recognition.
  2002;5(1):67-85.

\bibitem{roesch1999snort}
Roesch M, et~al.
\newblock Snort: Lightweight intrusion detection for networks.
\newblock In: Lisa. vol.~99; 1999. p. 229-38.

\bibitem{hopcroft2001introduction}
Hopcroft JE, Motwani R, Ullman JD.
\newblock Introduction to automata theory, languages, and computation.
\newblock Acm Sigact News. 2001;32(1):60-5.

\bibitem{wulf1995hitting}
Wulf WA, McKee SA.
\newblock Hitting the memory wall: implications of the obvious.
\newblock ACM SIGARCH computer architecture news. 1995;23(1):20-4.

\bibitem{wang2016overview}
Wang K, Angstadt K, Bo C, Brunelle N, Sadredini E, Tracy T, et~al.
\newblock An overview of micron's automata processor.
\newblock In: Proceedings of the Eleventh IEEE/ACM/IFIP International
  Conference on Hardware/Software Codesign and System Synthesis; 2016. p. 1-3.

\bibitem{tracy2016towards}
Tracy T, Fu Y, Roy I, Jonas E, Glendenning P.
\newblock Towards machine learning on the automata processor.
\newblock In: International Conference on High Performance Computing. Springer;
  2016. p. 200-18.

\bibitem{rabin1959finite}
Rabin MO, Scott D.
\newblock Finite automata and their decision problems.
\newblock IBM journal of research and development. 1959;3(2):114-25.

\bibitem{yu2006fast}
Yu F, Chen Z, Diao Y, Lakshman TV, Katz RH.
\newblock Fast and memory-efficient regular expression matching for deep packet
  inspection.
\newblock In: Proceedings of the 2006 ACM/IEEE symposium on Architecture for
  networking and communications systems; 2006. p. 93-102.

\bibitem{becchi2009evaluating}
Becchi M, Wiseman C, Crowley P.
\newblock Evaluating regular expression matching engines on network and general
  purpose processors.
\newblock In: Proceedings of the 5th ACM/IEEE Symposium on Architectures for
  Networking and Communications Systems; 2009. p. 30-9.

\bibitem{hopcroft1971n}
Hopcroft J.
\newblock An n log n algorithm for minimizing states in a finite automaton.
\newblock In: Theory of machines and computations. Elsevier; 1971. p. 189-96.

\bibitem{brzozowski1962canonical}
Brzozowski JA.
\newblock Canonical regular expressions and minimal state graphs for definite
  events.
\newblock Mathematical theory of Automata. 1962;12(6):529-61.

\bibitem{tabakov2005experimental}
Tabakov D, Vardi MY.
\newblock Experimental evaluation of classical automata constructions.
\newblock In: International Conference on Logic for Programming Artificial
  Intelligence and Reasoning. Springer; 2005. p. 396-411.

\bibitem{paxson1999bro}
Paxson V.
\newblock Bro: a system for detecting network intruders in real-time.
\newblock Computer networks. 1999;31(23-24):2435-63.

\bibitem{liu2011fast}
Liu C, Wu J.
\newblock Fast deep packet inspection with a dual finite automata.
\newblock IEEE Transactions on Computers. 2011;62(2):310-21.

\bibitem{wang2019hyperscan}
Wang X, Hong Y, Chang H, Park K, Langdale G, Hu J, et~al.
\newblock Hyperscan: a fast multi-pattern regex matcher for modern cpus.
\newblock In: 16th $\{$USENIX$\}$ Symposium on Networked Systems Design and
  Implementation ($\{$NSDI$\}$ 19); 2019. p. 631-48.

\bibitem{becchi2007hybrid}
Becchi M, Crowley P.
\newblock A hybrid finite automaton for practical deep packet inspection.
\newblock In: Proceedings of the 2007 ACM CoNEXT conference; 2007. p. 1-12.

\bibitem{kumar2007curing}
Kumar S, Chandrasekaran B, Turner J, Varghese G.
\newblock Curing regular expressions matching algorithms from insomnia,
  amnesia, and acalculia.
\newblock In: Proceedings of the 3rd ACM/IEEE Symposium on Architecture for
  networking and communications systems; 2007. p. 155-64.

\bibitem{jiang1993minimal}
Jiang T, Ravikumar B.
\newblock Minimal NFA problems are hard.
\newblock SIAM Journal on Computing. 1993;22(6):1117-41.

\bibitem{floyd1980compilation}
Floyd RW, Ullman JD.
\newblock The compilation of regular expressions into integrated circuits.
\newblock In: 21st Annual Symposium on Foundations of Computer Science (sfcs
  1980). IEEE; 1980. p. 260-9.

\bibitem{kumar2006algorithms}
Kumar S, Dharmapurikar S, Yu F, Crowley P, Turner J.
\newblock Algorithms to accelerate multiple regular expressions matching for
  deep packet inspection.
\newblock ACM SIGCOMM Computer Communication Review. 2006;36(4):339-50.

\bibitem{yang2011high}
Yang YH, Prasanna V.
\newblock High-performance and compact architecture for regular expression
  matching on FPGA.
\newblock IEEE Transactions on Computers. 2011;61(7):1013-25.

\bibitem{DFAGE}
Angstadt K, Wadden J, Dang V, Xie T, Kramp D, Weimer W, et~al.
\newblock MNCaRT: An Open-Source, Multi-Architecture Automata-Processing
  Research and Execution Ecosystem.
\newblock IEEE Computer Architecture Letters. 2018;17(1):84-7.

\bibitem{liu2020gpus}
Liu H, Pai S, Jog A.
\newblock Why gpus are slow at executing NFAs and how to make them faster.
\newblock In: Proceedings of the Twenty-Fifth International Conference on
  Architectural Support for Programming Languages and Operating Systems; 2020.
  p. 251-65.

\bibitem{wadden2018automatazoo}
Wadden J, Tracy T, Sadredini E, Wu L, Bo C, Du J, et~al.
\newblock AutomataZoo: A modern automata processing benchmark suite.
\newblock In: 2018 IEEE International Symposium on Workload Characterization
  (IISWC). IEEE; 2018. p. 13-24.

\bibitem{wadden2016vasim}
Wadden J, Skadron K.
\newblock VASim: An open virtual automata simulator for automata processing
  application and architecture research.
\newblock University of Virginia, Tech Rep CS2016-03. 2016.

\bibitem{hsdb}
Hyperscan-Compiling Patterns; 2020-08-07.
\newblock Available from:
  \url{http://intel.github.io/hyperscan/dev-reference/compilation.html#compilation}.

\bibitem{hscompile}
hscompile; 2020-08-07.
\newblock Available from: \url{https://github.com/kevinaangstadt/hscompile}.

\bibitem{hsscan}
Hyperscan-Scanning for Patterns; 2020-08-07.
\newblock Available from:
  \url{http://intel.github.io/hyperscan/dev-reference/runtime.html#runtime}.

\end{thebibliography}
\end{document}